\documentclass[prc,twocolumn,superscriptaddress,showpacs,showkeys]{revtex4}
\usepackage{graphics,epsfig}
\newcommand{\lsim}{\stackrel{\scriptstyle <}{\phantom{}_{\sim}}}
\newcommand{\gsim}{\stackrel{\scriptstyle >}{\phantom{}_{\sim}}}
\def\be{\begin{eqnarray}}
\def\ee{\end{eqnarray}}
\def\om{\omega}
\def\prt{\partial}
\begin{document}
\title{Resonance states below pion-nucleon threshold\\ and their
consequences for nuclear systems}
\author{Evgeni E.\ Kolomeitsev}
\affiliation{ECT$^*$, Villa Tambosi, I-38050 Villazzano  (TN),
INFN G.C. Trento, Italy }
\author{Dmitri N.\ Voskresensky}
\affiliation{Moscow Engineering Physical Institute, Kashirskoe shosse 31,
115409 Moscow, Russia} \affiliation{GSI, Planck Str.\ 1, D-64291
Darmstadt, Germany}

\begin{abstract}
Regular sequences of narrow peaks have been observed in the
missing mass spectra in the reactions $pp\rightarrow p\pi^+ X$ and $pd
\rightarrow ppX_1$  below pion-production threshold. They are
interpreted in the literature as  manifestations of supernarrow
light dibaryons, or  nucleon  resonances, or light pions forming
resonance states with the nucleon in its ground state. We discuss
how existence of such exotic states would affect properties of
nuclear systems. We show that the neutron star structure is
drastically changed in all three cases. We find that in the
presence of dibaryons or nucleon resonances the
maximal possible  mass of a neutron star would  be smaller  than the observational limit.
Presence of light pions does not contradict  the observed neutron star
masses. Light pions allow for the existence of extended nuclear objects
of arbitrary size, bound by strong and electromagnetic forces.

\end{abstract}
\maketitle

\section{Introduction}

The experimental search for exotic states in one-baryon and
two-baryon spectra  has a long history, see
review~\cite{Troyan93}. The enthusiasm has been  revived in the field
after two recent experimental reports~\cite{T97,Filkov01}. In
ref.~\cite{T97} the reaction $pp\rightarrow p\pi^+\, X$ was
investigated and three narrow peaks (width $\sim 5$~MeV) in the
missing mass spectrum were  seen at $M_X=1004$\,, $1044$ and
$1094$ MeV with high statistical significance.
Ref.~\cite{Filkov01} reported about the study of the reaction $pd
\rightarrow p\,p\,X_1$. Three peaks of the 5~MeV width were
clearly observed in the  missing mass $M_{p X_1}$ spectrum at
$M_{pX_1}=1904\pm 2$\,, $1926\pm 2$ and $1942\pm 2$~MeV. In the
missing mass $M_{X_1}$ spectrum the peaks are located at
$M_{X_1}=966\pm 2$\,, $986\pm 2$ and $1003$~MeV.
The correspondence between peaks in refs.~\cite{T97,Filkov01} was
discussed in~\cite{Beck01}.

The evident
regularity of the  peaks observed in~\cite{T97,Tat02} contrasts with previous
experimental data~\cite{Troyan93} limited by a lower resolution in
energy and lower statistics, although the compiled analyses of all
available data of ref.~\cite{Tat99,Tat02} shows that a similar pattern
glimpses in the old data too.

The true nature of these resonances is  obscure so far and three
interpretations are suggested in the literature: (i)~The peaks can
be assigned to supernarrow dibaryon resonances ($D'$), which have
been predicted theoretically within the bag~\cite{bag} and the
chiral soliton~\cite{Kopel95} models; (ii)~The peaks can be
interpreted as new nucleon resonances ($N'$); (iii) ~They are
bound states of one or two nucleons with several light
pseudoscalar particles ($\widetilde\pi$) having mass
$m_{\widetilde{\pi}}\simeq 21\pm 2.6$~MeV~\cite{W01}.

The above interpretations imply that the new particles should have an
exotic internal quark structure, having only a small overlap with
usual nucleon and pion states. Otherwise they would be produced
with a large probability and would manifest themselves in many
nuclear reaction channels, where they are not seen, as we know.
For the dibaryonic interpretation, the smallness of the  production cross
sections~\cite{Filkov01,Filkov00}  and estimates~\cite{Filkov00}
suggest that the pion-$D'$-deuteron coupling should be at least
ten times weaker than the standard $\pi N N$ coupling. Nucleon
resonances below the $\pi N$ threshold should have tiny couplings
to $\pi N$ and $\gamma N$ states otherwise they would spoil the
dispersion relations for $\pi N$ and Compton scattering, which
are fulfilled with good precision~\cite{disprel}. Smallness of
the $\gamma N N'$ coupling requires also smallness of $\om N N'$
and $\rho^0 N N'$ couplings. The coupling of "light pions"
($\widetilde\pi$) to usual pions and nucleons should also be
strongly suppressed in order not to contradict the data on $\pi\pi$ and
$\pi N$ interactions, cf.  discussions in ref.~\cite{W01}.

Despite the fact that the exotic states do not show up on the typical time scale
of strong interactions, according to their properties mentioned
above, they could manifest themselves in long living nuclear
systems, such as atomic nuclei and neutron stars (NS). They could be also
produced with a detectable probability in particle-nucleus
and nucleus-nucleus collisions as well as other rare probes
(photons, di-leptons, strange particles, etc). The small
probability of an elementary reaction is enhanced in latter case due
to a large number of interactions.

In the present notes we shall  discuss possible consequences of
the very existence of these three
suggested hypothetical states ( $D'$, ${N'}$, $\widetilde{\pi}$)
for the nuclear systems. In order to make our analysis more transparent, we
consider the lightest state  in the dibaryon spectrum,
$m_{D'}= 1904$~MeV, and the two lightest states in the nucleon
spectrum $m_{N'_1}= 966$~MeV and $m_{N'_2}= 986$~MeV.
For "light pions" we take $m_{\widetilde \pi}= 22$~MeV.

We shall model  the equation of state (EoS) of nuclear matter using the
parameterization~\cite{HH99}, which is a good fit to the optimal EoS
of the Urbana--Argonne group~\cite{APR98} up to a 4 times nuclear
saturation density, $\rho_0=0.16$~fm$^{-3}$ smoothly
incorporating the causality limit at higher densities. The energy
density, counted from the nucleon mass, is
\be\nonumber
E_N(\rho_N) &=& \rho_N\,\mathcal{E}_N(n=\rho_N/\rho_0,
x=\rho_p/\rho_N)\,,
\\ \label{apr}
\mathcal{E}_N(n,x) &=& \mathcal{E}_0 n \frac{2.2-n}{1+0.2 n}+
S_0\,n^{0.6}\,(1-2\,x)^2\,, \ee where $\rho_{p(n)}$ is the proton
(neutron) density, $\rho_N=\rho_p+\rho_n$,
$\mathcal{E}_0=-15.8$~MeV and $S_0=32$~MeV.  Having the
compressibility modulus $K\simeq 200$~MeV, this EoS allows for the
NS masses up to $2.2 M_\odot$\,, where $M_\odot=2\times 10^{33}$~g
is the solar mass.

\section{Light dibaryons}
Consider a mixture
of nucleons and dibaryons with the total baryon density $\rho_B$ and the
dibaryon density  $\rho_{D'}$.
The energy density of such a system is given by
\begin{eqnarray} \label{en-d}
E= E_N(\rho_B-2\,\rho_D)+\delta m_{D'} \rho_{D'}
+E^{\rm pot}_{D' N}+E^{\rm pot}_{D' D'},
\end{eqnarray}
where    $\delta m_{D'} =m_{D'}-2\,m_N = 28$~MeV, and
quantities $E^{\rm pot}_{D' N}$ and $E^{\rm pot}_{D' D'}$
are potential-energy densities of dibaryon-nucleon and
dibaryon-dibaryon interactions.
Following \cite{Tat99} we assume here that
dibaryons are composed of neutrons (having isospin zero),
obey Bose statistics, and occupy, therefore, single lowest-energy state.
They are compact systems. Thus we may assume that they do not undergo
the Mott transition in the baryon matter at densities we are interested
in.

Consider first the case of a small  dibaryon concentration.
Expanding (\ref{en-d}) in $\rho_{D'}$ for isospin symmetrical matter and
$\rho_{D'}\ll\rho_B$, we obtain
\begin{eqnarray}\label{en-dexp}
E \approx E_N (\rho_B ) -(2\,\mu_N^{(0)} - \mu_{D'}^{(0)} )\rho_{{D'}}+
\kappa\rho_{{D'}}^2/2\,.
\end{eqnarray}
Here $\mu_{D'}^{(0)}=\delta m_{D'}+\delta\mu_{{D'}N}$ and
\begin{eqnarray}\label{cor}
&&\mu_N^{(0)} =\left.\frac
{\partial E_N}{\partial \rho_{B}}\right|_{\rho_{B}=\rho_N}\,,
\,\,\,\delta \mu_{{D'}N} =\left.\frac
{\partial E^{\rm pot}_{{D'} N}}{\partial \rho_{{D'}}}\right|_{\rho_{{D'}}=0},
\nonumber \\ \label{kappa}
&&\kappa =4\left.
\frac{\partial^2 E_N}{\partial
\rho_{B}^2}\right|_{\rho_{B}=\rho_N}
+\left.\frac{\partial^2( E^{\rm pot}_{{D'} N}+
E^{\rm pot}_{{D'} {D'}})}{\partial
\rho_{{D'}}^2}\right|_{\rho_{{D'}}=0}\!\!\!.
\end{eqnarray}
Minimizing (\ref{en-dexp}) with respect to $\rho_{{D'}}$,
we find that for $\kappa>0$ a
dibaryon admixture becomes energetically favorable, if
 $2\,\mu_N>\mu_{D'}$. Then  the dibaryon density and the  energy density
gain are:
\be\label{dib}
\rho_{{D'}}\simeq\frac{2\,\mu_N^{(0)}-\mu_{D'}^{(0)}}{\kappa}\,, \quad
\Delta E_{D'}\simeq
- \frac{1}{2\kappa}\Big(2\,\mu_N^{(0)}-\mu_{D'}^{(0)}\Big)^2\,.
\ee
To model the unknown dibaryon-nucleon interaction  we assume
that ${D'}N$ potential is proportional  to $NN$ potential and
$\delta\mu_{{D'}N}=\zeta (\mu_N^{(0)}-\epsilon_{\rm F})$.
Here $\epsilon_{\rm F}$ is the Fermi energy of the isospin symmetrical
matter, $\epsilon_{\rm F}=\epsilon_{{\rm F},p}=\epsilon_{{\rm F},n}$,
with $\epsilon_{{\rm F},i}= ({m_N^2+p_{{\rm
F},i}^2})^{1/2}-m_N$ and  $p_{{\rm
F},i}=(3\,\pi^2\rho_i)^{1/3}$ for $i=n,p$.
The exotic internal structure of light dibaryon states implies that the parameter
$\zeta$ is small.
Specifically, ref.~\cite{Filkov01,Filkov00} put a constraint $0.1\gsim \zeta\geq
0$. A naive quark counting would give, on the other hand, $\zeta=2$ (such an assumption
is analogous to that one in ref.~\cite{FBKM98}).
A repulsive  dibaryon self-interaction can be introduced  with
$E_{{D'}{D'}}^{\rm pot}=\pi\,f_{{D'}{D'}}\,\rho_{D'}^2/m_{D'}$\,,
where $f_{{D'}{D'}}>0$ is the  ${D'}{D'}$ scattering length. It
can be estimated using the assumption, justified in the bag
model~\cite{bag}, that the hard core radius in the third power
scales as the particle mass, $f_{{D'}{D'}}\sim r_{N}^{\rm cor}\,
(m_{D'}/m_N)^{1/3}\sim 0.4$~fm, with $r_N^{\rm cor}\sim 0.3$~fm.
The coefficient $\kappa$ can be evaluated  using that at saturation
density
$\prt^2 E_N/\prt \rho_N^2|_{\rho_N=\rho_0}=K/(9\rho_0)$,  and
assuming
that $E_{{D'}N}^{\rm pot}$ is exhausted by the term $\propto\rho_{D'}\,
\rho_N$, giving , thereby,
$\prt^2 E_{{D'}N}^{\rm pot}/\prt \rho_{D'}^2|_{\rho_{D'}=0}=0$.
Then we have $\kappa\simeq 610$~${\rm MeV} \cdot {\rm fm}^{-3}$\,.

Applying now the results (\ref{dib}) to atomic nuclei we find that for
the favorite choice  $\zeta\simeq 0$ we have $\mu_{D'}^{(0)}>0$
and dibaryons are absent  in atomic nuclei,  since nucleons are
bound with $\mu_N^{(0)}<0$. Thus, \emph{ the dibaryon
interpretation of experiments~\cite{T97,Filkov01,Tat99} does not
contradict the atomic nucleus experiments}, if their interaction
with nucleons is, indeed, sufficiently weak. Note, that  choosing
$\zeta=2$, according to a quark counting, we would have at
saturation density $2\mu_N^{(0)}-\mu_{D'}^{(0)}\approx p_{\rm
F}^2/m_N-\delta m_{D'} \approx 46$~MeV corresponding to $\rho_{D'}
\sim 0.5\rho_N$. This would evidently contradict the known
properties of nuclei. The critical value of $\zeta$ at which
dibaryons would just appear in nuclei is $\zeta_c=(\delta
m_{D'}-2\, \mu_N^{(0)})/(\epsilon_F-\mu_N^{(0)})$\,, that gives
$\zeta_c\simeq 0.8$ for $^{56}$Fe with $\mu_N^{(0)}\simeq -8$~MeV.

We turn now to NS, which could in principle contain an  admixture of
dibaryons. Consider a NS consisting
of neutrons, protons, dibaryons, electrons, and muons.
Following the  assumptions above
we model the nucleon-dibaryon potential energy as:
\be\label{edn}
E_{{D'}N}^{\rm pot}=\frac{\zeta}{2}\, (V_n+V_p)\,
\rho_{D'}\,,
\quad
V_i=\frac{\prt E_N}{\prt \rho_i}-\epsilon_{{\rm F}, i}\,,
\ee
with $i=n,p$.
The chemical potentials of nucleons and dibaryons are given by:
\be\nonumber
\mu_{i}=\frac{\prt (E_N+E_{{D'}N}^{\rm pot})}{\prt \rho_{i}}
\,,\, \mu_{D'}=\delta m_{D'}+
\frac{\prt (E_{D'N}^{\rm pot}+E_{{D'}D'}^{\rm pot})}{\prt
\rho_{D'}}\,.
\ee
The composition of the NS matter is controlled by  conditions of
charge-neutrality  and $\beta$-equilibrium~\cite{G97}.
The latter  means, particularly, that $\mu_{D'}=2\mu_n$.
Calculations (in line with  the standard procedure~\cite{G97})
show that the presence of dibaryons would completely change  the NS composition.
Dibaryons appear at very small densities. For homogeneous baryon
matter it happens at $\rho_B\simeq 0.2 \rho_0$.
With further increase  of the baryon density
they expel protons
and strongly reduce the neutron concentration, e.g. at
$\rho_B\simeq 4.5 \rho_0$ we have $\rho_n/\rho_B\approx
\rho_D/\rho_B\simeq 0.33$ and $\rho_p/\rho_B\simeq 10^{-4}$\,.

The total energy density, $E_{\rm tot}$, is given by (\ref{en-d})
plus the kinetic energy of the leptons. The latter contributions are
very small since the number of leptons, being equal to the number
of protons, is suppressed in presence of dibaryons. The pressure
is equal to $P=\sum_a \mu_a\, \rho_a-E_{\rm tot}$, where the sum
goes over all species present in the system.

\begin{figure}
\begin{center}
\includegraphics[width=6cm,clip=true]{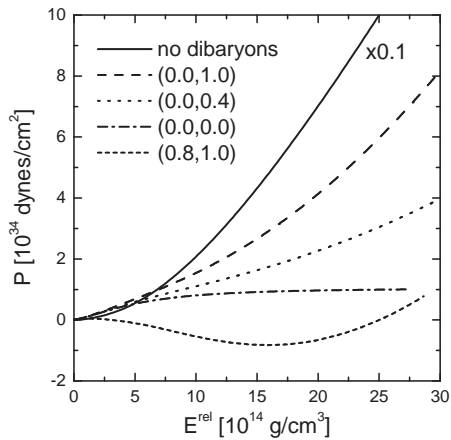}
\end{center}
\caption{
Pressure  of the NS matter as a function of the energy density
with and without dibaryons.
Numbers in brackets indicate parameter sets ($\zeta$,  $f_{D'D'}$ in fm).
The curve for the dibaryon-free case is scaled by factor 0.1.
}
\label{fig:dibar}
\end{figure}

In Fig.~\ref{fig:dibar}  we show the pressure vs. the full
relativistic energy density, $E^{\rm rel}= m_N \rho_B +E$, of the
NS matter with and without dibaryons, calculated for  various
values  of  parameters $\zeta$ and $f_{D'D'}$\,. We observe that
dibaryons make the EoS very soft.  In presence of dibaryons the
pressure   is typically smaller by an order of magnitude than
that one  without dibaryons.  The stiffness of EoS is increased
with increase of $f_{D'D'}$\,. However, even for the case
$f_{D'D'}=1$~fm, corresponding to the most stiff EoS among those
presented in Fig.~\ref{fig:dibar}, the maximum mass of the NS, as
we calculated, is significantly less than the observational lower
limit $1.4M_\odot$. Moreover, an increase of $f_{D'D'}$ up to
unrealistically large values $f_{D'D'}\sim 10$~fm would  not lead
to the increase of the NS  maximum mass up to $1.4M_\odot$. EoS
becomes softer if one assigns dibaryons to the isospin one states,
assuming thereby the  existence of charged dibaryons. The same
happens for finite values of  $\zeta$. For $\zeta=0.8$ and
$f_{D'D'}= 1$~fm, the pressure would be negative in some density
interval, indicating that the system would undergo the collapse to
a superdense state, which would be bound even in the absence of
the gravity.

Thus, the analysis above, allows us to conclude that
\emph{ the existence of the light dibaryons
would be in severe contradiction with information on NS
masses}.

\section{Nucleon resonances}
We turn now to another possible interpretation of
experiments~\cite{T97,Filkov01} with the help of exotic nucleon
resonances  of isospin 1/2. Consider two lightest resonances
$N'_1$  and $N'_2$ with $\delta m_{N'_1}=m_{N'_1}-m_N= 28$~MeV and
$\delta m_{N'_2}=m_{N'_2}-m_N= 48$~MeV. We denote neutral and
positively charged particles as  $n'_{1,2}$ and $p'_{1,2}$ by
analogy with neutrons and protons. Our treatment of the $N'$
resonances in the NS matter is similar to that one for hyperons,
cf.~\cite{G97}. However, the critical density for the appearance
of $N'$ is much smaller due to the smaller $N'$ mass. In line with
ref.~\cite{T97,Tat02}, we assume that the $N'$ resonances have the
quark structure quite different from the nucleon one, and,
therefore, the Pauli principle is not operating between them and
nucleons.

The energy density of the system with the baryon density
$\rho_B$, composed of nucleons and ${N'_{1,2}}$ resonances with the density
$\rho_{N'_{1,2}}=\rho_{n'_{1,2}}+\rho_{p'_{1,2}}
$ is:
\begin{eqnarray}\nonumber
E &=& E_N (\rho_B-\rho_{N'_1}-\rho_{N'_2})+\sum_{i'=p'_{1,2}, n'_{1,2}}
E^{\rm kin}_{i'}
\\  \label{en-n}
&+&E^{\rm pot}_{{N'}N}+ E^{\rm pot}_{{N'} {N'}}\,,
\end{eqnarray}
where $E^{\rm kin}_{p'(n')}$ denotes the kinetic energy of the
$p'(n')$ resonance counted from  the nucleon mass.
Similarly  to (\ref{en-d}), the
quantity $E_{N'N(N'N')}^{\rm pot}$ stands for the $N'N (N'N')$
potential-energy density.
Assuming $\rho_{{N'_1}}\ll\rho_B$ ($\rho_{N'_2}=0$ for this case) we   expand
(\ref{en-n}) in $\rho_{{N'}}$ and  obtain for the case of
isospin  symmetrical matter
\begin{eqnarray}\nonumber
&&E=E_N (\rho_B ) -(\mu_N^{(0)} -\mu_{N'_1}^{(0)})\, \rho_{N'_1}+
\frac35 \frac{\rho_{N'_1}^{5/3}}{m_{N'_1}}\,(\frac32\,\pi^2)^{2/3}\,,
\\ \label{en-nexp}
&& \mu_{N'_1}^{(0)}=\delta m_{N'_1}+{\prt E^{\rm pot}_{N'N}}/{\prt
\rho_{N'}}\big|_{\rho_{N'_1}=\rho_{N'_2}=0}\,,
\end{eqnarray}
where we retained $\rho_{{N'_1}}^{5/3}$ terms but dropped the higher order terms,
and $\mu_N^{(0)}$ is defined, as in (\ref{en-dexp}).
Minimization of the energy density over the $\rho_{{N'_1}}$ yields
the following $N'$ density and energy gain:
\begin{eqnarray}
\nonumber
\rho_{{N'_1}}&\simeq& 2\,[m_{N'_1}\,(\mu_N^{(0)}-\mu_{N'_1}^{(0)})]^{3/2}/(3\,\pi^2)\,,\\
\label{r-n}
\Delta E_{N'_1} &\simeq&-\frac{4}{15\,\pi^2} \,m_{N'_1}^{3/2}\,(\mu_N^{(0)}
-\mu_{N'_1}^{(0)})^{5/2}\,,
\end{eqnarray}
if $\mu_N^{(0)}>\mu_{N'_1}^{(0)}$ and zero otherwise.
As for dibaryons, we express the $N'_1$ chemical potential through the
nucleon chemical potential,
$\mu_{N'_1}^{(0)}=\delta m_{N'_1}+\zeta'\,(\mu_N^{(0)} - \epsilon_{\rm F})$\,,
with the same constraint $0.1\gsim \zeta' \geq
0$. For $\zeta'=0$ there are no $N'$ resonances in atomic nuclei, since
$\mu_{N'}^{(0)}=\delta m_{N'}>0$ and $\mu_N^{(0)}<0$.
A  quark counting  suggests $\zeta'=1$, and, correspondingly, we would get
$\mu_N^{(0)}-\mu_{N'_1}^{(0)}
\approx p_{\rm F}^2/(2\,m_N)-\delta m_{N'_1}
\approx 9$~MeV in atomic nuclei. This gives $\rho_{N'}\simeq 0.04\,\rho_0$ and
$\Delta E_{N'}\simeq 0.2$~MeV, which are rather small contributions.
The critical value of $\zeta'$, when $N'_1$ resonances  could just appear
in atomic nuclei, is $\zeta'_c \simeq 0.9$.
For $0.1\lsim \zeta' \lsim\zeta'_c$, even if $N'$ resonances   are
absent in atomic nuclei, they could be produced and detected
in  nuclear reactions. This is, however, not the case.
Therefore, $\zeta'$ should be small and for small $\zeta'$
\emph{the existence of light nucleon resonances would not
contradict known properties of atomic nuclei}.

Consider now the NS matter in $\beta$-equilibrium consisting of nucleons,
$N'_1$, $N'_2$ and leptons.
We express the $N'N$ and $N'N'$ potential energies  as:
\be\nonumber
E_{N'N}^{\rm pot}&=&\zeta'\,\left[V_n\,(\rho_{n'_1}+\rho_{n'_2})+
V_p\,(\rho_{p'_1}+\rho_{p'_2})\right]\,,
\\ \nonumber
E_{N'N'}^{\rm pot}&=&\frac{\pi
f_{N'N'}}{m_{N'_1}}(\rho_{N'_1}+\rho_{N'_2})^2\,,
\ee
where the nucleon potentials, $V_{n,p}$ are given in
(\ref{edn}), and we take $f_{N'N'}\sim r_N^{\rm cor}\simeq
0.3$~fm. The chemical potentials of nucleons and $N'$ are  then as follows
\be\nonumber
\mu_{i}=\frac{\prt (E_N+E_{N'N}^{\rm pot})}{\prt
\rho_{i}} \,,\,\,
\mu_{i'}=\epsilon_{{\rm F},i'} +\frac{\prt
(E_{N'N}^{\rm pot}+E_{N'N'}^{\rm pot})}{\prt \rho_{i'}}\,,
\ee
where $i=n,p$, $i'=n'_{1,2},p'_{1,2}$ and
$\epsilon_{{\rm F},i'}=(m_{N'_{1,2}}^2+p_{{\rm F}, i'}^2)^{1/2}-m_N$ with
$p_{{\rm F},i}=(3\,\pi^2\,\rho_{i'})^{1/3}$.
The $\beta$-equilibrium  requires that $\mu_n=\mu_{n'_1}=\mu_{n'_2}$ and
$\mu_p=\mu_{p'_1}=\mu_{p'_2}$.

\begin{figure}
\begin{center}
\includegraphics[width=8.3cm,clip=true]{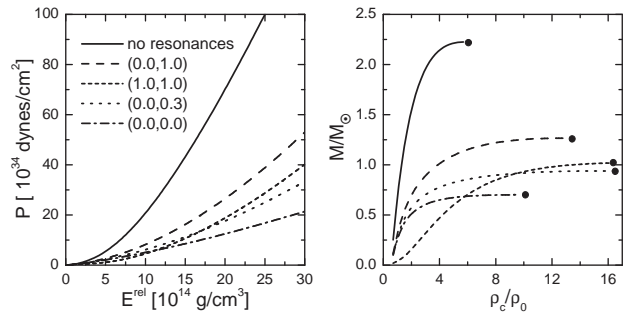}
\end{center}
\caption{
Left panel: Pressure as a function of the relativistic energy density
of the NS matter with and without $N'_{1,2}$ resonances. The curves are presented
for different parameter sets ($\zeta'$, $f_{N'N'}$ in fm).
Right panel: The NS mass as a function of the central baryon density. The line styles
are identical in both panels. }
\label{fig:nres}
\end{figure}

The composition of the NS matter would be significantly changed in
presence of $N'$ resonances. For $\zeta'=0$ and $f_{N'N'}=0.3$~fm neutron and proton
concentrations decrease continuously starting from $\rho_B \simeq
0.2\rho_0$ (in the approximation of  homogeneous medium)
with neutrons, being replaced, first, by $n'_1$ and then after $1.2\rho_0$
by $n'_2$. At $\rho_B\simeq 4\rho_0$ the
chemical potential $\mu_{p'}$ exceeds $m_{N'}$. At this point all protons are
replaced  by $p'_1$ resonances, whose population grows slowly,
reaching 0.5\% at $10\rho_0$. The $p'_2$ resonances do not appear.
The pressure  of the NS matter with $N'$ resonances
is shown in Fig.~\ref{fig:nres} (left panel) as a function of the energy
density $E^{\rm rel}$.
We see that $N'$ resonances make the EoS softer, decreasing the pressure
by factor about 2 to 5 for $0\le \zeta'\le 1$ and $0.3~{\rm fm}\le
f_{N'N'}\le 1~{\rm fm}$.
The softness of EoS increases with increase of $\zeta$ and decrease of
$f_{N'N'}$.
In the right panel of Fig.~\ref{fig:nres} we show the masses of NS
cores as a function of central baryon densities, $\rho_c$.
The full dots indicate the maximum masses of NS, $M_{\rm max}$.
Since astronomical observations put the limit
$M_{\rm max}\geq 1.4 M_\odot$~\cite{G97},
we can  conclude  from Fig.~\ref{fig:nres}
that \emph{the existence of $N'$ resonances
would  contradict information on the
NS masses for $\zeta'\ge 0$ and $f_{N'N'}\lsim 1$~fm.}
To reach the observational limit of $1.4M_\odot$ the $N'N'$ repulsion  should
be as large  as with $f_{N'N'}\gsim 2$~fm.
A further stiffening of the EoS of nuclear matter (1)
would not help since
then $N'$ resonances would appear at even smaller densities,
leading to the same consequences for NS.

\section{Light pions}
We follow now the assumption of
ref.~\cite{W01} that there exist
light pions ($\widetilde{\pi}$) whose
strong coupling constant is essentially suppressed.

Our approach here is similar to that one used for pion condensate systems, cf.
\cite{M71,MSTV90} and refs. therein.
Negative light pions could  be accumulated in nuclei due to
reaction  $n\rightarrow p +\widetilde{\pi}^-$,
if $\mu_n -\mu_p\gsim\mu_{\widetilde{\pi}}$\,.
Having a suppressed strong coupling constant, light negative pions
satisfy the Klein-Gordon equation
\begin{eqnarray}\label{kg}
\Delta \phi +[(\omega -V)^2 -m_{\widetilde{\pi}}^2 ]\,\phi  =0,
\end{eqnarray}
where $\phi$ is the $\widetilde{\pi}^-$
wave function and
$V$ is the Coulomb potential well.
Multiplying (\ref{kg}) by $\phi^{*}$ and averaging over the volume
we find the  solution
\begin{eqnarray}\label{kg-sol}
\omega =\overline{V}+\big(m_{\widetilde{\pi}}^2 +
\overline{k^2}+(\overline{V})^2 -\overline{V^2}\big)^{1/2},
\end{eqnarray}
where $\overline{k^2}=\int |\nabla \phi |^2 d\vec{r}$.
The critical value
of the pion energy $\omega$ at the $\widetilde{\pi}^-$
condensation point is
$\omega =\mu_n -\mu_p \simeq 0$, for  isospin-symmetrical nuclei.
For the light pions the Compton wave length is $\sim 10$~fm
that is larger than the radius ($R$) of the nucleus.
Thus, $\overline{k^2}$ is the dominating term in
(\ref{kg-sol}).

Let us first model a nucleus as
a spherical potential well  of the constant
depth $V=-V_0\simeq -Z e^2/R$ for $r<R\simeq 1.2 A^{1/3}$~fm and $V=0$ for
$r>R$, where $e$ is the electron charge and $A$ is an atomic number.
Then the pion spectrum follows from the equation
$kR\cot(kR)=-\lambda R$, with
$k=\sqrt{(\omega +V_0)^2 -m_{\widetilde{\pi}}^2 }$ and $\lambda =
\sqrt{m_{\widetilde{\pi}}^2 -\omega^2}$.
In the limit $Rm_{\widetilde{\pi}} \ll 1$
the spectrum of deeply bound states
can be found from the condition $\cos(kR )=0$
and in the opposite limit, $Rm_{\widetilde{\pi}} \gg 1$, from $\sin(kR )=0$.
In the most favored case for the $\widetilde{\pi}^-$
condensation ($\cos(kR )=0$)
the critical condition reads
$V_0= (m_{\widetilde{\pi}}^2 +{\pi^2}/(4{R^2}))^{1/2}$\,.
For atomic nuclei (with the atomic number $A\lsim 200$) we then estimate
$(m_{\widetilde{\pi}}^2 +{\pi^2}/(4{R^2}))^{1/2}>50$~MeV,
and the critical condition is not yet achieved. Solution of (\ref{kg-sol})
with the realistic Coulomb potential does not change this conclusion.
\emph{Thus, there is no $\widetilde{\pi}^-$ condensate in
atomic nuclei,} their size and the value of the Coulomb field
are  too small for that.

In refs.~\cite{MSTV90,V77} it was argued that,
if there existed light bosons of the mass $<30$~MeV, then
there could  exist exotic objects, "nuclei-stars",
of an arbitrary size,  having the density $\rho \sim \rho_0$, and
being bound by strong and electromagnetic interactions.
The light pion proposed in ref.~\cite{W01} could play such a role.
To demonstrate this idea, let us consider
the spherical nuclear ball of a constant density $\simeq \rho_0$ and the
radius  $1~{\rm km}\gg R\gg m_{\widetilde{\pi}}^{-1}$,
consisting of $A$ nucleons. Then we may neglect both surface and gravitational effects.
The total energy is given by
\begin{eqnarray}
{\cal E} \simeq \mathcal{E}_0\, A +S_0\, (A-2\,Z)^2 /A +Z\,m_{\widetilde{\pi}}.
\end{eqnarray}
Minimizing the energy over
$Z$ at fixed $A$, we get $Z/A \simeq 0.41$ and the corresponding binding energy
${\cal E}_N /A\simeq -5.7$~MeV. Although
this quantity is slightly larger than
the binding energy per particle for the $\alpha$ particle
($-7$~MeV),
the $\widetilde{\pi}$-nucleus cannot decay into $\alpha$
particles and light pions or atomic nuclei and light pions.
The total energy per particle  of $\alpha+2\,\widetilde{\pi}$ is $\ge-7~{\rm
MeV}+m_{\widetilde{\pi}}/2=3$~MeV. This is larger than the
energy per particle of the initial bound state ($\simeq -5.7$~MeV).
(This, however, does not exclude  a possibility of the weak radioactive
decays of $\widetilde{\pi}$-nuclei.)
At finite temperature (excitation energy) $T>T_{c1} \sim$ several MeV
the binding energy reaches zero and
condensate melts. The system, however, still exists as a whole, being in
a metastable state, until excess of  energy is smaller than $Z
m_{\widetilde{\pi}}$.
Such an excited object cools down via $\gamma$- and
$\nu$-radiation from  weak $\widetilde{\pi}$ decays.
At $T>T_c \sim m_{\widetilde{\pi}}$ the system decays,  expanding into the
vacuum.

\begin{figure}
\begin{center}
\includegraphics[width=8.3cm,clip=true]{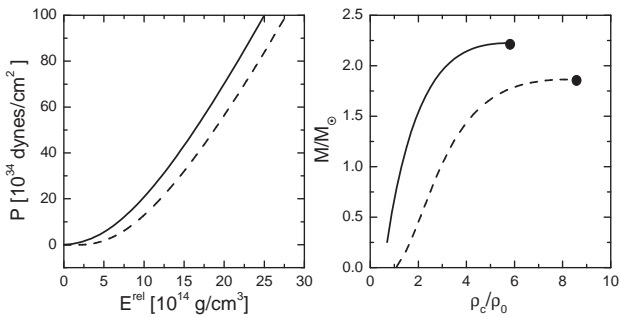}
\end{center}
\caption{ Left panel: Pressure of the
NS matter as a function of the relativistic energy density
without (solid line) and with (dashed line) the presence of light
pions $\widetilde\pi$. Right panel: The NS mass as a
function of the central baryon density.
The line styles are the same in both panels.} \label{fig:lpions}
\end{figure}

Consider light pions in NS matter.
Here we first assume that the $N'$ resonance and  $D'$ dibaryon
do not exist as elementary particles,
and the experiments~\cite{T97,Filkov01,Tat02} are explained
with the help of the usual nucleons and the light pions only.
In case of homogeneous baryon matter
$\widetilde{\pi}^-$ mesons could  appear already at the density
$\rho_B \simeq 0.05\rho_0$,
when the electron chemical potential reached $m_{\widetilde{\pi}}$.
Then at higher densities, in the NS interior,  $\mu_e=m_{\widetilde{\pi}}$,
and the positive proton charge is compensated by the negative charge of
$\widetilde{\pi}^-$ meson condensate.
It leads, first, to the isospin equilibration of NS
matter. E.g. at $\rho=2\rho_0$ the ratio $\rho_p/\rho_B\simeq 0.44$,
increasing  further  with the baryon density. Second, with the
$\widetilde{\pi}^-$ condensate the EoS of NS matter
becomes softer, as it is illustrated in Fig.~\ref{fig:lpions}.
The pressure is negative at  $0.3\rho_0<\rho_B <\rho_0$,
indicating that  the system is self-bound
at $\rho_B = \rho_0$ (when $P=0$) even in the absence of the gravity.
The softening of the EoS reduces the maximal mass of the NS but still keeps this value
above the observational limit of $1.4~M_\odot$.
\emph{Thus the existence of light pions would not contradict  the
observed NS masses.}
If light pions coexist with N' resonances and/or D' baryons  the EoS
 would be softer than in the case without light pions. Such a situation
 would contradict to observable neutron star masses.

In heavy-ion collisions, $\widetilde{\pi}$ mesons would
contribute to dilepton spectra and could be seen as a peak
in the dilepton spectrum at $\sim 20$~MeV energy.

\vspace*{5mm}

\section{Summary}
Being motivated by recent experimental studies
\cite{T97,Filkov01,Tat99} and their interpretations, we have
investigated how the existence of exotic light dibaryon, nucleon
resonance and pion states, would manifest itself in nuclear
systems. We have shown that dibaryons and $N'$ resonances below
the  $\pi N$ threshold would be absent in atomic nuclei, if their
interactions with nucleons are sufficiently small, as it  is
demanded by their production rates. Also light pions  cannot be
accumulated in atomic nuclei.

In the neutron star matter the new exotic states could manifest
themselves remarkably: they would drastically change the composition of a
neutron star and make the  equation of state much softer.
The equation of state with  dibaryons and nucleon resonances  becomes so soft that
it would not be able to support neutron stars  with the observed masses.
Thus the dibaryon and nucleon-resonance interpretations of the above
mentioned experiments should be questioned.

The existence of light pions would  not lead to a contradiction with observed  neutron
star masses. The presence of  light pions would allow the existence of
abnormal nuclei ($A\gsim 10^3$) and ``nuclei-stars'' of arbitrary size,
bound by strong and electromagnetic interactions.

Thus, the importance of astrophysical consequences of the low-mass  resonance states
should strongly motivate further experimental investigations,
as well as  a search for new theoretical interpretations.

\begin{acknowledgments}
The authors thank W.~Weise and L.V.~Filkov for helpful remarks.
D.N.V. acknowledges  hospitality and support of GSI Darmstadt.
His work has been supported in part by DFG (project 436 Rus 113/558/0),
and by RFBR grant NNIO-00-02-04012.
\end{acknowledgments}

\end{document}